# Macroscopic Continuous Approach versus Discrete Approach, Fluctuations, criticality and SOC.

## A state of the question based on articles in Powders & Grains 2001

### P. Evesque
Lab MSSMat, UMR 8579 CNRS, Ecole Centrale Paris
92295 CHATENAY-MALABRY, France, e-mail: evesque@mssmat.ecp.fr

**Abstract:**

*The macroscopic continuous approach of the mechanics of granular media assumes that the whole system of discrete variables (contact locations, contact forces,…) can be replaced by continuous field equations relating stress $\sigma$ and strain $\varepsilon$ on macroscopic scale. On the contrary, recent approaches contest this validity on the basis that microscopic studies show the existence of large fluctuations of forces, of chains of forces… This paper tries and establishes the state of this question using recent works reported at Powders & Grains 2001 which have studied the microscopic variables, their fluctuations and their evolution. This paper shows that these results validate the macroscopic approach despite the existence of these fluctuations. It concludes that the representative elementary volume is of few grains most of the time, except in some peculiar cases .*

**Pacs # :** 5.40 ; 45.70 ; 62.20 ; 83.70.Fn

___

In the mid 60s, the pioneering work of Dantu [1] has demonstrated the heterogeneous nature of the force distribution in a granular assembly under static conditions. This result was confirmed rapidly. The scientific community was surprised and has urged a series of works in the domain of the mechanics of granular matter in order to determine whether the continuous modelling was remaining pertinent or was invalid. So, the mid 70s have seen a careful re-examination of many macroscopic mechanical behaviours on granular matter and sand. This study has concluded (more or less) that this local force heterogeneity was not invalidating the continuous approach as far as the size of the granular sample was large enough to contain many grains. This led also to introduce the notion of the Representative Elementary Volume (REV). However, it was still lacking at that time a complete understanding, allowing the exact prediction and calculation of the macroscopic properties from the microscopic ones. Furthermore, as technology was less powerful than nowadays, there were still missing a series of works aimed at elucidating the properties of a granular assembly, of its local mechanics, *i.e.* at the scale of few grains, and of its statistics. It was also needed to get a clear definition of what shall be considered as the REV size, how it varies under experimental conditions and so on.

Meanwhile, physics has developed a new series of concepts such as dilation symmetry, fractals, renormalisation group (RG), frustration, chaos and strange attractor in order to describe the physics of second order phase transition, of disordered electric conductors, of disordered matter, of glasses, of complex dynamical systems, of hydrodynamics, of weather forecast,… This became possible mainly owing to the development of computer technology, which has given access to more and more complicated simulations and has allowed treating more and more complex information and data. This new technology has renewed completely the goals and the possibilities of investigation in many domains of physics.

In the domain of granular matter, this had some impact as soon as the early 80s: Using the technological break-through of computer, Cundall and Strack [2] have developed their Discrete Element Method (DEM) of simulating the mechanical behaviour of granular matter in order to get a deeper scientific insight. Slightly later (1985), Roux, Stauffer & Herrmann [3] have simulated the force distribution in a 2d assembly with small disorder; they have shown that this disorder propagates on large distances and that fractals was needed to describe it, breaking up the concept of simple averaging.

In the same way, Back, Tang and Wiesenfeld (BTW) [4] have proposed a simple computer algorithm, *i.e.* the BTW model, to study avalanching of grains at the free surface of a computer-made sandpile; it has been found that this algorithm generates spontaneously avalanches of all sizes, *i.e.* 1 grain,…, few grains,…, N grains, …, with a distribution decreasing as a power law $1/N^\alpha$. This led to the concept of Self-Organised Criticality (SOC), which led in turn to think that granular matter was the archetype of critical behaviour. This idea has been probably reinforced by the fact (i) that Bak and co-workers have tentatively applied SOC to understand the physics of seism (because seisms exhibits a multiple-scale mechanism) and (ii) that seism mechanics could be the archetype of Earth mechanics, hence of soil mechanics and hence of granular matter.

So, this has contributed to develop in some part of the scientific community the idea that granular matter was a complex highly-heterogeneous system, which exhibits spontaneously multiple hierarchical scales. This took place in the late 80s and in the 90s. This has cast a serious doubt on the validity of the continuous approach and stimulated a tremendous series of works within these last few years either aimed at determining the domain of validity of the continuous approach, or to understand why the continuum mechanics can/cannot be applied. Assuming so, physicists were passing over the work done in the 70s by the mechanics specialists, which has concluded to the validity of the macroscopic approach. Anyway, this doubt is indeed an important and good motivation to produce a series of new studies at local scale and to fill up the existing lack of data.

So, most recent papers in physics of granular matter is aimed at invastigating complexity: Is the mechanics of these media predominantly dominated by highly non linear effect, with spin glass and strange attractor as archetypes? This is all right: good goal ! But we shall expect in turn that the conclusion of these articles comes back to this question and tries to conclude: Do these new works strengthen the existence of such an anomalous behaviour? Or do they strengthen the classic approach and the classic mechanics viewpoint?

Indeed, this lack of conclusion bothers me. Hence, the goal of this paper is just to introduce this discussion. It will use *Powders & Grains 2001* Proceedings, since this meeting has tried to settle the state on the research on granular media. This will allow to show that the continuous approach is probably a good approximation as far as granular samples remain of the size used in labs and as far as special working points are excluded. (Indeed, as the main goal is the identification of an "always anomalous" behaviour, whose possible archetypes are SOC or glassy systems, the existence of few anomalous working points has no consequence on the general behaviour, and hence on the discussion, as recalled in the appendix).

So the paper tries to define the actual status of few main questions asked in the physics literature: *Question 1:* Does granular matter exhibit a quasi-static rheological law which can be defined at a scale of few thousands grains? (The answer will be Yes from [5] as soon as grains exhibit solid friction). *Question 2:* Is it needed to distinguish between two networks of forces? (The Question has been stated in [6]; the answer will be No after discussion using [5] and [6]; however an exception will be given in question 3). *Question 3:* Are two sub-networks needed in the case of size segregation? (The answer will be Yes, after discussion using [6] and [7]). *Question 4:* Is critical behaviour observed in granular matter? (The answer will be Yes but only at some special working point, from [8]). *Question 5* will be divided in two parts. *Question 5a:* how stress propagates from a point perturbation? (Answer: Using classic equation often); And *Question 5b:* Can the stress dip in conic granular sandpile be explained within classic approach? (Answer: yes). *Question 6:* Does ultrasonic wave propagate in granular matter? (The answer is yes for large wavelength $\lambda$; when $\lambda$ becomes of the order of the grain size, scattering and diffusion become efficient). *Question 7:* How to define the representative elementary volume? *Question 8 and Conclusion:* What is the REV of a granular medium? (The answer is few grains).

Few questions of interest will not be treated: As we are only interested in looking for SOC behaviour, we will not discuss papers which find an intermediate scale, such as in [9]: since this length is well defined and corresponds to few grains only, it cannot generate criticality by itself. This is also why we will not discuss papers which are aimed to describe (i) regular macroscopic behaviours, (ii) a way to perform the micro-macro passage using some mean field approach or approximation; this includes in particular the papers describing anisotropic behaviours only.

In the same way, the problem of avalanching is out of the scope of the paper since it is well accepted now that it does not produce SOC easily. Indeed, it has been shown that (i) quasi-static classical rheology, with solid friction and pressure-dependent dilatancy, would predict a surface flow with critical fluctuations in some special cases [10], (ii) that experimental avalanching obeys these classical laws at least partly [10-12], (iii) that "critical" fluctuations have been observed in small systems [12,13] and predicted from classical rheology [10]. However, there are still few problems to solve in avalanching which are: which complexity of rheology is really needed [14]? Is dynamic friction always negligible [12]?

## 1. Question 1: On the definition of a stress-strain rheological law [15, 16]:

*Frictionless grains :* In his thesis Combe [15, 16] has investigated the deformation of 2d packings of frictionless disks under a uniaxial compression. He has found that the deformation occurs via a series of reorganistion of the contact distribution; the curve characterising the stress-strain path, in the system of coordinates {deformation $\varepsilon_1$, relative deviatoric stress $q=(\sigma_1-\sigma_3)/\sigma_3$)}, exhibits a staircase shape with random steps alternatively vertical and horizontal, with variable jumps $\delta q$ and $\delta\varepsilon_1$. Each vertical step corresponds to loading without deformation and each horizontal step to deformation and to a change of contact distribution. The statistics of $\delta q$ and $\delta\varepsilon$ have been determined for samples of various sizes. In the case of disordered packing, it has been found that the $\delta q$ statistics exhibits an exponential shape, *ie.* $P(\delta q) \approx \underline{\delta q} \exp(-\delta q/\underline{\delta q})$, with a mean $\underline{\delta q}$ which varies with the number N of grains of the sample according to $\underline{\delta q}(N) \propto \underline{\delta q_o} N^{-\alpha}$, with $\alpha=1.13\pm0.06$ and $\underline{\delta q_o}= 4.37\pm0.06$. However the statistics of deformation jumps $\delta\varepsilon_1$ is much wider and does not accept a mean; for instance, for large $\delta\varepsilon_1$, the distribution scales as $P(\delta\varepsilon_1) \propto (\delta\varepsilon_1)^{-(1+\mu)}$, with $\mu=0.53\pm0.05$; for small $\delta\varepsilon_1$, there is a cut-off $\varepsilon_o$ which ensures the convergence of the probability distribution; this cut-off scales with the number of grains as $N^\beta$ with $\beta=2.11\pm0.02$.

This result is remarkable, since it demonstrates that no relation of the kind $\underline{\delta q}=f(\underline{\delta\varepsilon_1})$ exists in the limit of large systems in the case of frictionless grains; in other words it demonstrates the absence of a rheological law in the case of frictionless granular material, which seems to contradict the common mechanical approach proposed by mechanics. A question arises then: is this result general or only valid in the case of the present study of frictionless grains?

*Grains with solid friction :* In their article of *Powders & Grains 2001* [5], Combe & Roux have continued their study, examining the case of grains with solid friction.

They have found in the limit of rigid grains, that the q *vs.* $\varepsilon_1$ curves are still made of a succession of vertical and horizontal increments of random amplitude; but their statistics accept both a mean $\underline{\delta q}$ and $\underline{\delta\varepsilon_1}$. And $\underline{\delta q}$ and $\underline{\delta\varepsilon_1}$ decrease when the number N of grains of the sample increases. This demonstrates that the rheological law $\underline{\delta q}=f(\underline{\delta\varepsilon_1})$ does exist in this case even if the curve is made of steps, since it becomes more and more regular as N increases.

As the grains used are not rigid, the earliest stage of the deformation exhibits a stress-strain law which depends on the contact law; but, the case of rigid grains, with its staircase shape, is recovered as soon as the strain becomes large enough.

As the means $\underline{\delta q}$ and $\underline{\delta\varepsilon_1}$ decrease when the system size grows, all these results are compatible with older conclusions from the mechanics community, hence confirming and strengthening them; it is indeed in agreement with classical experimental results of soil mechanics.

There is still a big difference with the classic theoretical understanding : in the classic view, it was thought that deformation occurs via the combination of two distinct processes ; the first one is a slow continuous evolution of the geometry of the contact- and of the force- distributions; the second one, which occurs randomly, is a rapid change of the contact network, which occurs due to the breaking of old contacts and the creation of new contacts ; this last process may provoke a dynamical instability and a discontinuous change of the force network. In this classic approach, one neglects quite often the second process [17, 18]. On the contrary, the numerical study of Roux and Combe lets think that the first process, of a continuous adaptive deformation, does not occur.

*Remarks:* At this stage it is worth noting few points :

(i)  As the evolution of the mechanical properties of the assembly seems to be controlled mainly by dynamical effects and by the evolution during non equilibrium periods, this explains perhaps why dynamical algorithm such as [19] gives behaviours similar to quasi-static ones such as [2].

(ii) Hence, the word quasi-static is likely inappropriate, since deformation occurs during a succession of microscopic "stick-slip events", whose slips are dynamical events allowing discontinuous restructuration of the contact network. So, the fact that classical rheology calls this regime quasi-static is probably confusing; however, it is just because the speed of these contact restructuration is much faster than the macroscopic speed of deformation, so that this last one is independent of the time needed for the fast events. In turn, the fact that simulation with friction gives realistic behaviours, *i.e.* similar to classic experimental results, implies likely that the chronology of the micro-slips and of the contact changes is not so important; this may prove that the possible evolution at each stage is manyfold; so the system has to choose between many new possible configurations, which are able to accommodate the new stress field when unbalance occurs; each set of configuration is a completion and evolution chooses randomly between these completions; as they are numerous, evolution chooses always among the most probable completions.

(iii) Also, in the simulations of Combe & Roux [5], or of Jean or of Moreau [19], dynamics effects are not taken into account completely, but partially only, since these algorithms either search for only one of the possible new stable configurations or use only one of the possible collision chronologies. Due to this, these algorithms do not solve completely the mathematical problem and do not find the exact solution which satisfies the precise initial condition; but they find just a plausible trajectory which looks realistic. But, it turns out that this method allows good prediction as soon as one uses contacts with solid friction. In turn, this demonstrates that the mean behaviour is not sensitive to the real details of the dynamics in this case. This is an effect of statistics in large systems, which is used for instance in Monte Carlo algorithm: Indeed, the chosen combination is always one of the most probable evolutions; this is also such an effect which makes valid the law PV=nRT of gases. So the results of Combe & Roux and of Jean or Moreau with friction contacts tend to demonstrate that the granular system behaves in a "regular" way, which means that correct averaging can be defined as usual in the limit of large systems with many internal degrees of freedoms, the internal variables being here either the force network and/or the velocity of the grains. However, this result is no more true when frictionless grains are used, since correct q *vs.* $\varepsilon_1$ curve cannot be defined there.

(iv) One shall note also that the simulations in [5] exhibit only a series of continuous increase of q (δq is always positive); Combe & Roux tell that it is not a general trend and that some negative-δq increments can be/have been observed. But they have not given any counterexample with negative δq. On the other hand, they do give examples of positive and negative values of volume variations, *i.e.* see Fig. 4.30 on p. 147 of [15] .

(v) We turn now to energy consideration on frictionless grains. We call perfect plasticity the mode of deformation when no volume change is observed in mean; this deformation mode shall be observed spontaneously at large deformation, since an assembly of grains in contact cannot expand or contract too much, *i.e.* to infinity. General theorem on work gives that the energy furnished to the system is $\delta W = \sigma_3(q\delta\varepsilon_1/\sigma_3 + \delta\varepsilon_v) = \sigma_3\delta\varepsilon_1 (q/\sigma_3 - K)$ , where $K = -\partial\varepsilon_v/\partial\varepsilon_1$ is the dilatancy. δW shall be equal to the energy dissipated during the deformation. We examine now what this rule implies:

We first consider the case of frictionless grains. Owing to the frictionless nature of the contact law, the system cannot dissipate energy so that its perfect-plasticity regime shall be characterised by q=0 in mean, which imposes that the friction angle φ=0 . (By passing, this general result contradicts some assertion of Cambou [20], who has considered the possibility of having φ≠0 in the case of frictionless grains; this demonstrates in turn, if it was needed, that the micro↔macro passage proposed in [20] is not complete). Anyhow, this work equation, $\delta W = \sigma_3(q\delta\varepsilon_1/\sigma_3 + \delta\varepsilon_v)$, shows that only dilatant material of frictionless grain can be stabilised in quasi-static when q>0.

Coming back to Combe & Roux simulations, Fig. 4.20 of [15] shows (a) that the studied system exhibits often a dilatant dynamical evolution, (b) that it can exhibit a contractant dynamical evolution, but in this case, it is associated often with a large dynamical deformation δε. Indeed, points (a) and (b) seem coherent with what global energy rule can predict, *i.e.* instability for contracting evolution.

Another question arises also: as contractant event with q positive shall generate kinetic energy, this kinetic energy shall be dissipated before next equilibrium is achieved. In experiment, this is done via dissipation at contacts; in molecular dynamics simulation with frictionless grains, this is done via damping. On the contrary, in the case of Combe & Roux simulations, since the grains are frictionless, this is not required explicitly; but it is assumed implicitly via the algorithm that searches for a new static equilibrium with a new configuration of contacts when equilibrium is disrupted.

This has probably some consequences: at each step, an equilibrium configuration is looked for; this one requires a new dilatant configuration because q is positive and since $\delta W = (q\delta\varepsilon_1 + \sigma_3\delta\varepsilon_v)$. So, this may explain why configurations that would exhibit little dilatancy are not found and why configurations exhibiting large dilatancy are obtained systematically, because they ensure stability; in turn this may explain the staircase shape of the evolution.

Anyhow, the simulation assumes also a "strong dynamic damping". This ensures the stability of a contractant dynamic evolution ; hence it explains why contractant steps are observed and perhaps. But we shall note also that the larger strain steps are produced during contractant steps. Is this due to the damping used?

(vi) We turn now to the case of assemblies with solid friction. In this case, Combe and Roux simulations lead to similar behaviours, characterised by an intermittent evolution of stress then strain. Whether this contact network can be considered as a rigid then fragile structure seems to be a general trend of Combe & Roux simulations. It means also that dissipation is never due to pure friction, but shall contain also some dynamical damping. This has some implication at the level of the rheological law: If these simulations are able to reproduce classical rheological behaviour, they shall also reproduce the Rowe's law of dilatancy [17]. In this case, it would mean that the explanation proposed by Rowe himself is not correct, since it is based on a quasi-static analysis of losses due to solid friction during slidings.

(vii) Isostaticity: It is known also that a system of frictionless grains which can deform is isostatic; this implies that the complete system of forces is determined when halve the forces imposed from the exterior are known [18]; the other halve being determined via the set of relations which ensures local equilibrium of all grains [18]. So, as static equilibrium shall be ensured at each step, it shall satisfy $(q\delta\varepsilon_1+\sigma_3\delta\varepsilon_v)=0$, or $\delta\varepsilon_v/\delta\varepsilon_1=-q/\sigma_3$, in the case of frictionless grains, which demonstrates (i) that a volume variation shall exist during the quasi-static deformation, (ii) that the dilatancy $K=-\partial\varepsilon_v/\partial\varepsilon_1$ depends on q. This is not observed in Combe & Roux paper.

As a conclusion on Combe & Roux works, we may associate SOC behaviour to the case of frictionless grain, because no rheological behaviour can be defined in this case and because it generates a broad distribution of $\delta\varepsilon$, with a power law tail. This example is then quite interesting for a pedagogical purpose, because it allows to explain what kind of trouble one can observe when no good averaging can be done. However, the main result of ref. [5] is that a regular stress-strain behaviour is recovered as soon as some solid friction is introduced in the contact law, which demonstrates that assemblies of "classic", *i.e.* dissipating, grains exhibit regular trends.

## 2. Question 2 : On strong and weak sub-networks : [6]

The paper of Radjai [6] studies the force transmission in 2D or 3D granular media by looking at the statistical distribution of the contact forces and at their correlation during quasi-static shearing. It recalls that such transmission shows up a correlation length of 10-grain diameters about. It finds in particular that the contact network can be viewed as the sum of two disjoined sub-networks $\mathcal{R}_w$ & $\mathcal{R}_s$, called respectively the sub-network of weak contacts ($\mathcal{R}_w$) and the one of strong contacts ($\mathcal{R}_s$); these sub-networks have two different distributions $P_w(F)$ and $P_s(F)$ of contact forces F: *i.e.* labelling $F_1$ and $F_2$ the force component respectively normal to- and tangent at- the contact surface, one has $P_w(F_i)= k_i (F_i/\langle F_i\rangle)^{\alpha i}$ for $\mathcal{R}_w$, with $F_i<\langle F_i\rangle$, and $P_s(F_i)=k_i\,e^{\beta i(1-F_i/\langle F_i\rangle)}$ for $\mathcal{R}_s$, with $F>\langle F\rangle$, where $\langle F\rangle$ is the mean force and where $k_i$, $\alpha_i$ & $\beta_i$ are 3 constants which satisfy $1/k_i=1/(1-\alpha_i)+1/\beta$ and $\beta_i^2=(1-\alpha_i)(2-\alpha_i)$. Simulations show (a) that 60% of contacts have a force below $\langle F_i \rangle$, (b) that $\alpha_1$ varies from 0 to 0.3 during shear, (c) that $\alpha_2=0.3$ always, (d) that the contribution of the weaker set of forces has an isotropic contribution in mean, (e) that $R_w$ contributes to 29% about of the isotropic stress p (p is defined as p=Trace($\underline{\underline{s}}$)/d, where d is the space dimension, *i.e.* d=2 in 2 dimensions and d=3 in 3 dimensions). The second network, $R_s$, corresponds to contacts supporting the largest forces; it is the one which supports the deviatoric stress, *i.e.* shear; its contribution is then anisotropic.

The article [6] tends to argue that the separation into two distinct sub-networks is an important fact which shall be taken into account to describe the physics of the system. This is this point which is needed to discuss.

In order to settle the problem and demonstrate that this assertion is not obvious, let us first remind the case of a classic gas with few kinds of atoms, *i.e.* $O_2$, $N_2$ or He, in concentration $n_1$, $n_2$, …. Indeed, we know that the macroscopic law pV=nRT is valid for this mixture, with $n=(n_1+n_2+…)$. But looking at its microscopy, the speed distribution will be the sum of few different laws $\exp[-m_i v^2/(2kT)]$, because each molecule of mass $m_i$ has its own mean velocity. Hence, each kind of molecule should pertain to a specific sub-network owing to the above approach. However, one knows also that these sub-networks define a single and same phase because they are in statistical interaction, in equilibrium and occupy the same volume. By the way, this is ensured because none of them carries a specific meaning since all these statistics (and their moments) converge. Indeed, as far as no chemical reaction is involved, one knows that the existence of few kinds of molecule do not perturb the macroscopic law pV=nRT, which applies equivalently whether the gas contains a single class of atoms or few. This is due to the intermixing of the different components. (Remark: some care has to be taken however, because some varaiables may depend on the exact nature of each compound; this is the case of the specific heat and of the sound speed of a gas which depend on the polyatomicity; but this does not make the gas multiphasic).

We can now come back to the case of granular media. The strong network $R_s$ has a simple exponential statistics which ensures the convergence of its different moments. The weak network $R_w$ is power-law distributed, *i.e.* $P_w(F_i)=k_i (F_i/\langle F_i \rangle)^{\alpha_i}$; but it has a cut-off at large forces, *i.e.* $F_i < \langle F_i \rangle$; and since $\alpha_i$ is positive, this ensures also the convergence of any moment of physical interest (positive order) ; better, the convergence is always ensured even when one does not know the exact number of contacts $N_w$ because this number $N_w$ cannot be infinite in any finite sample and because it scales proportionally to the number of particles the sample contains. On the other hand, it is possible that the two networks play a different part: for instance, the strong network might wear the stress , whereas the weak one might stabilise the contact distribution. However, this does not imply that they do not pertain to the same phase ; on the contrary, if they pertain to two distinct phases, this would imply that the two sub-networks will be frozen . This is what will be studied in the following sub-section:

***Statistical mechanics approach:*** Simulations show that the two sub-networks $R_s$ and $R_w$ are sensitive to any tiny modification: (i) changing the calculus precision modifies the contact forces and the repartition of the contacts in between the two sub-networks. (ii) Changing also slightly the value of a single force, at a given stage of deformation, modifies the force distribution and changes also the repartition of the contacts in between the two sub-networks. Nevertheless, statistical characteristics of each sub-network distribution remains unchanged. These results allow to conclude that the two networks are in equilibrium and pertain to the same phase:

These results demonstrate that different force configurations exist which satisfy similar external conditions, indicating that the system is hyperstatic most likely. Since the repartition of contacts in between the two sub-networks is sensitive to the calculus precision or to some tiny modification of a single force, each contact has some probability to pertain to a set or to the other, and this membership varies with time randomly; hence a mean field can be applied and a single phase can be defined. Furthermore as each sub-network contain many contacts and is in equilibrium with the other one their statistical characteristics shall be well defined; this explains why the distribution characteristics of each sub-networks do not depend on calculus precision and do not change when some tiny force is changed slightly: the two systems obey statistical conditions of many-body systems and hence to well defined "equilibrium" relation.

This needs to be confirmed by further studies; but it lets already foresee that any possible configuration can be chosen at random at well. It lets then think that Monte-Carlo technique can be used. Indeed, this is already used in some other algorithms such as [19] where the order of collisions is not computed exactly; and these algorithms give quite good behaviours. Hence this strengthens the existence of a single phase. This may also justify some ergodicity, within some restriction.

Indeed Fig. 6 of [6] confirms the analysis, at least partly: it gives the rate X of contact redistribution from one sub-network as a function of the strain $\varepsilon$ and shows that this rate X of redistribution is at least 0.1% per $\delta\varepsilon=10^{-6}$ with bursts 20 times faster. So, if one assumes that the contact interchange is a process independent of $\varepsilon$ (in mean), one finds that the complete regeneration of the strong-network needs $\Delta\varepsilon\approx10^{-3}$ ; this $\Delta\varepsilon$ remains quite small. One can check the independence of X by counting the total number N of contacts which has changed; it shall vary as $N=\int X\, d\varepsilon$ , with X being given by Fig. 6 of [6]; and N shall grow linearly with $\varepsilon$.

So, as a conclusion, the simulations [6] tend to prove that any contact pertain to a sub-network, with probability p and to the other one with probability (1-p). Its membership is not frozen but varies randomly with deformation and/or with other perturbation ; the structure of each sub-network is correlated at small scale and may exhibit anisotropy. Nevertheless, when the grains are identical, the status of each contact is fixed at random and evolves rapidly with deformation or with applied perturbation so that the strong- and the weak- sub-networks are permanently intermixed and hence form a single phase [22]. This conclusion contradicts the one of ref. [6]; but it will be strengthened in the next section by the new interpretation of the results of Antony & Ghadiri [7] on segregation.

It shall be noted however, that the characteristics of this unique phase can be complex and its definition may require many parameters which describe the different quantities of interest, *i.e.* the fabric tensor, the mean density of contacts, the mean density of contact supporting the shear (or pertaining to $R_s$) ……….

At last, it is worth noting that the present conclusion is in agreement with the facts reported in §-1 and with §-1 conclusions since simulations of Combe & Roux have shown that the force network is sensitive to details and that a stress-strain law can be defined correctly as far as the grains exhibit friction.

## 3. Question 3 : On segregation during quasi-statics [7]

When a larger grain is inserted into a granular medium of smaller grains, the simulations reported in [7] demonstrate that the large particle support always less deviatoric stress than the surrounding medium, *cf.* Fig. 2 of [7]. (However its stress field is not hydrostatic [7]).

Nevertheless, let us first consider for a while that this larger grain supports exactly a hydrostatic pressure. Then, transposed into the model of the previous section, *i.e.* §-2, this result implies that the contacts pertaining to the larger grain pertain always to the weak sub-network. Conversely, it implies also that these contacts cannot jump freely from one sub-network to the other one. This breaks the validity of a single-phase approach and forces the contact pertaining to a large grain to pertain to a well defined sub-network, which generates locally a new phase. In other terms, this imposes that the two sub-networks are frozen in the neighbourhood of a larger grain, which imposes the existence of two distinct frozen phases; as these two phases do not support the same stress they shall evolve differently; this is then the main motor of the segregation process.

So, results of refs. [5-7] can be integrated in a single scheme of interpretation; this one is strengthened hence. Furthermore, it is reinforced by experimental results, which demonstrate the existence of phase separation due to size segregation.

The fact that size segregation is a well-known phenomenon which forces the larger grains to separate spontaneously demonstrates that the two distinct networks introduced by Radjai in [6] have an important meaning exclusively when grains are different. Otherwise the two sub-networks are intermixed permanently and form a single phase.

Conversely, experiments show they are intermixed with identical grains; otherwise one should observe locally two counter-flows with constant direction, the direction being imposed by the stress tensor. Indeed, this proves that the two-phase interpretation of [6] has no meaning in the case investigated in [6].

In the case investigated in [7], the existence of two sub-networks is needed; however their definition is slightly different from the one defined in [6], since the larger grains support some small but non zero deviatoric stress. The weak sub-lattice which has to be defined actually for describing segregation [7] corresponds to a cut in the contact force distribution which is different from the one uses in [6]; this cut defines two other mean stress tensors different from the one defined in [6].

Results of [6] demonstrate that a single large sphere has to be considered as making a single phase already, whose characteristics are different from the surrounding granular medium. This is why segregation occurs at once as soon as a single large grain is added, without accepting any dilution quota; this makes granular segregation different from precipitate of chemicals. Indeed, in size segregation of particles, there is no need of (large_grain-large_grain) interaction to produce a demixing where as A-A interaction is needed to get precipitate of A.

So this section has demonstrated that a discrimination into 2 different sub-lattices is required as soon as two grain sizes are involved : in this case, the larger grain condense one sub-network, imposing a local demixing into two distinct phases. This makes the system of segregation to look "similar" to the one of precipitation of chemicals; but, as segregation occurs with a single grain, the analogy runs with the case of precipitates having already a typical size. So, it falls partly in the domain of classic physics and chemistry, and not in the domain of SOC.

## 4. Question 4 : on critical behaviour in quasi-statics 2d Couette experiment [8]

It has been found [8] that the 2d Couette rheology of an assembly of photoelastic disks or pentagons exhibits some kind of critical behaviour when the density $\gamma$ of the medium decreases till $\gamma_c^+$. ($\gamma_c$ depends on the object shape and on its size distribution; for the disk experiment it was $\gamma_c=0.776$). For instance as $\gamma \to \gamma_c^+$, (i) the system is more compressible, (ii) the stress fluctuations become intermittent, (iii) the length $\underline{L}$ of the typical stress chain increases from 2 to 5 grain diameters, (iv) the statistics of chain length changes from exponential to broader (*cf.* Fig. 9 of [8]), (v) the mean stress $\underline{\sigma} \to 0$ as $\underline{\sigma} \propto \{(\gamma-\gamma_c)/\gamma_c\}^\alpha$, (but, as discussed further, $\alpha$ does not seems to be a critical exponent since $\alpha$ depends on the particle shape, *i.e.* $\alpha=4.04$ for disks $\alpha=2.15$ for pentagons. When $\gamma \to \gamma_c^+$, the distribution of the larger force F exhibits an exponential tail, *i.e.* exp(-F/F$_o$), whose typical value decreases when $\gamma \to \gamma_c^+$. Also, the force distribution changes; it becomes more linear near $\gamma_c^+$.

These results can be interpreted in the scheme of critical phenomena due to the "divergence" of physical quantities when approaching $\gamma_c$. So, $\gamma_c$ would be the critical point. In this case $\alpha$ shall be considered as a critical exponent. But one knows that critical exponent are constant parameter independent of many details. The fact that $\alpha$ varies with the grain shape would indicate that the nature of grain-grain interaction changes from pentagons to cylinders!?

Can we really assert that the above results depend only on the granular medium? For instance, what is the effect of the boundary? Do the results depend the statistics of the set-up imperfections? We cannot answer these questions yet.

Anyhow, the anomalous scaling, if it can be fully developed, seem to occur just at $\gamma_c$. And near $\gamma_c$, the anomalous behaviour develops only on restricted length scales, since no divergence is observed. Indeed, this excludes SOC unfortunately (that SOC can be generated in granular media when σ is small is an old idea [23], which turns to be "wrong").

Anyhow, it is possible to develop an other argument which can plead not only against SOC but also against critical behaviour ; it comes from the example of avalanching in real granular piles. Indeed, critical avalanching have been predicted for extended samples and found experimentally exclusively in finite small sandpile [12, 13] only. It means that experimental SOC requires σ≈0 **and** $L<L_o$. On the contrary, in larger sandpile surface flow, which means σ=0 or σ≈0 or σ>0 but large length, obeys a sub-critical bifurcation [10-12], at least in most cases [12]. A possible explanation for the "cross-over" from critical to sub-critical bifurcation can be explained in the case of avalanching by the fact that the macroscopic rheology requires a large REV, to be defined; so when the pile is small it exhibits fluctuations which look like SOC because they are induced by the grain structure, but when it is larger than $L_o$, its mechanics becomes governed by a macroscopic law which inhibits the critical behaviour and forces the subcritical one. As usual the cross-over between the two behaviours is obtained when the amplitude of the sub-critical process, *i.e.* the avalanche size Δθ, becomes equal to the uncertainty of the slope δθ=$D/L_o$, D being the grain size.

Such a scheme could work also in the case of Behringer's experiments [8] performed on larger samples: in larger systems, the medium could be in continuous contact with the boundary applying on it a given finite homogeneous stress or it could be in contact in some part of the boundary and not in the other part forcing a bistable quasi periodic motion; so it is possible that a more regular law be obtained in larger sample size, *i.e.* larger than the REV.

## 5. Question 5: On stress propagation and distribution

In recent years, the way stress propagates in granular media has been the subject of debate. It has been argued that (i) it could generate anomalous propagation [24], (ii) be the cause of some "fragile" effect [25], and (iii) can be the signature of the jamming transition [26]. This will discussed briefly in sections 5, 6.

### *5.1. Question 5.a: On propagation of stress from a perturbation at a point.*

It has been argued recently that stress in granular media should obey a hyperbolic differential equation in space, so that stress perturbation should propagate along lines in space [24]. This way of thinking was disputed with reason at *Powders & Grains 1997* [27], but controversy is still holding. So new 2d experiments has been performed recently by Behringer et al. [8] using photo-elastic discs and force perturbation located on some point. The work confirm mainly prior findings: when disordered packings are used, it allows concluding to the absence of "anomalous" stress propagation as soon as the length is larger than few grains. Moreover comparison with 3d results reported at Powders & Grains 2001 on more complicated situations [28] deserves to be done, because these results do exhibit important similarity and are explained viq classic approach; so it demonstrates the generality of classic behaviour, even in more complicated situations.

In ordered systems, "anomalous" propagation occurs on larger distance [8], but "linear" propagation vanishes as well at larger scale, *i.e.* 5 to 10-grain diameters [8].

This last conclusion has perhaps to be taken with caution in view of the work of Roux et al. [3] which has examined the case of 2d assembly with little disorder and proved that the effect of such a disorder propagates on quite large length scale. Anyhow, the ordered-, or little disordered- situations are not expected in natural assemblies; so they have not to retain our attention and shall not modify the conclusion of the experiments: no anomalous stress propagation is observed in 2d disordered assembly.

However, this is a surprising conclusion for a mechanics specialist: Indeed, classical behaviour occurs because stress propagates according to elliptic differential equation and "anomalous" propagation occurs with hyperbolic equations. So perhaps the actual problem is the reversed one:

One knows (i) that perfect plasticity was introduced in mechanics a long time ago to describe different cases, (ii) that hyperbolic equations govern stress in such materials and (iii) that granular materials can exhibit perfect plasticity. So the real question is: why is it difficult (impossible?) to produce systematic examples of anomalous stress propagation from punctual perturbation, if one excepts the few cases with regular grains of ordered assembly. This is surprising.

## *5.2. Question 5.b: On the stress dip in a conic pile*

An other topical problem in recent years has been the stress dip which is observed at the bottom of some conic piles: Can this dip be described with classical approach of continuum mechanics or not? Does the dip validate new approaches [27]? Indeed, the answers given in the Proceedings of *Powders & Grains 2001* by [29] are quite clear and they are in agreement with other works [30-32]. But the talk was slightly confused, and it seems better to come back briefly on the results.

Indeed, [29] reports on both experimental study and simulations using Discrete Element Method (DEM) of the stress distribution in a conic sandpile.

In what concerns the experiment of [29], it shows that the stress field, and its dip, is sensitive to many effects such as the building process, the size of the pile, the cohesion and the density profile. This confirms our own experimental results with a centrifuge [30] which have (i) invalidated the hypothesis of the so-called Radial Stress Field (RSF) scaling [30-31], (ii) which have proved the sensitivity of the stress distribution to the pile history and to the effective gravity [30].

DEM simulations of [29] reproduce indeed most of the features obtained in experiments. It is worth noting that these features are reproduced also by our own simulations [32] using Finite Element (EF) method associated with macroscopic constitutive law. Both simulations invalidate in particular the RSF scaling hypothesis; but both simulations also reproduce the stress field, its sensitivity to building history, to the exact rheological law, to the density.

So, this forces to conclude that DEM and EF methods give correct prediction; it means then that EF simulations, which are based on a continuous approach, allow to predict (at least approximately) the sandpile mechanics, when appropriate constitutive law is used. Hence this conclusion is in agreement with the previous partial conclusions of §-2, §-3 and §-4 and strengthens them and the validity of the continuum approach as soon as the pile size is large enough.

## 6. Question 6: sonic and ultrasonic wave propagation in granular media

It was asserted few years ago [33] that sound was not propagating normally in granular media. This result can astonish the mechanics and engineering community, which uses sound propagation in soil, sand,… to prospect oil in ground…

Anyway, this result [33] has been obtained on the basis of a too short study that has explored the propagation of sound on few-grain distances only; this sound was emitted by a single grain and detected via a single grain. Hence due to the smallness of the path, to the smallness of the emitter and detector, the experiment was extremely sensitive to fluctuations of the contact-force network, leading to giant fluctuations of the path length and of the sound speed.

Jia [34] has repeated similar experiment, but has used larger distances, and larger detector and emitter sizes. In this case he has got more reliable results. Indeed, Jia's work has demonstrated that sound propagates normally in granular media, as far as the wavelength $\lambda$ of the ultrasound is larger than few times the grain size D. When $\lambda$ becomes of the order of the grain size, one observes also efficient wave scattering and wave diffusion. This generates a speckle that can be observed and analysed using ultrasonic transducers. Indeed, measurement of the speckle intensity as a function of the transducer size follows normal law for averaging; measurement of sound velocity as a function of stress exhibit law similar to those already published at smaller frequency… This confirms, if it was needed, that classic approach of sound waves in soil and sand, is correct.

## 7. Question 7 : On Representative elementary Volume of solid, fluid and gas

In many recent publications the problem of fluctuations is related to the size of the representative elementary volume (REV); is this association correct? We will demonstrate it is not since REV requires that (i) mechanical coupling between adjacent REV exists and (ii) that fluctuations decrease with volume size as any Gaussian noise. So, the REV size may depend on the physical quantity of interest. Hence, what is the REV size $\xi^3$ of a granular medium.

When mechanics tries and defines REV, it means the REV which has to be used from the point of view of the mechanics of continuum media. In order to settle the question more correctly, let us start with more classic cases where we knows that continuum mechanics applies; we start and consider the case of a fluid (liquid or gas).

## *7.1. The REV of a classic gas*

In fluid mechanics, the application of the "action-reaction" principle requires the collisions of the molecules with one another or with a wall. Hence, the natural microscopic scale of continuum mechanics is the mean free path $l_c$, whose definition is the mean distance that a molecule travels in between two collisions with other molecules; when the fluid is made of identical spherical molecules of diameter D, of mass m at concentration n, the mean free path $\xi=l_c$ is given by :

$$n \, l_c \, (\pi D^2) = 1 \qquad (1)$$

Eq. (1) shows that $l_c$ diverges as n→0 or when D→0. Eq. (1) implies in particular that the number of molecules or atoms in the elementary volume $l_c^3$ scales as $N_{lc}=n(\pi D^2 n)^{-3}=n^{-2}\pi^{-3}D^{-6}$, which indicates that the total number of particles diverges as $n^{-2}$ when n→0 or $D^{-6}$ when D→0. The problem is now to determine whether this lengthscale $l_c=\xi$ is important or not, for the mechanics description.

A question arises: *What can be defined on volume smaller than $V_\xi=\xi^3=l_c^3$?* In this case, molecule-molecule collision occurs unlikely and the fluid obeys the dynamics of independent bodies. Anyway, within some uncertainty, one can still define the mean density n, the total number of particles $N=nL^3$, the mean pressure p, the temperature T, and the classical law of perfect gas $pL^3=nRT$. Furthermore, one can deform the sample with a shearing rate $\gamma = v_o/L$, by imposing a differential speed $v_o$ in between two parallel walls; this allows to measure the transfer of momenta from one wall to the other one due to successive collisions of molecules with the two walls; hence, this allows to define the shear stress as a function of the shear rate and the viscosity coefficient; (however, this requires that molecules get in thermal equilibrium with the wall after each collision with this wall). In the same way, changing the volume of the cell allows to define the compressibility of the gas.

Does this means that all mechanical characteristics of this gas can be observable. For instance, can one define a stress distribution, *etc.*, an acoustic propagation? In fact we will show in the next paragraphs that (i) one cannot define a diffusion process, (ii) one cannot define sound, nor sound propagation, (iii) one cannot define local thermodynamics transformation, except on the total volume.

**L<$l_c$=ξ ▶** *No molecular diffusion, the Knudsen regime:* Let us consider a sample of size $\Omega=L^3$ containing an ideal gas of point-like particles, *i.e.* D=0, of mass m at temperature T. In this case, $\xi=l_c$ becomes infinite, so that L<<ξ and the particles become independent; so, they move back and forth from one wall to the others, with a constant speed v during their free flights, changing of speed after each collision with the walls; the distribution of speed is the Boltzmann one with a mean speed $v_i$ in direction i (i=x,y or z) defined by its standard deviation $<v_i^2>=(2k_BT/m)$, with $k_B$ being the Boltzmann constant. A remarkable point is that molecules do not diffuse in such a gas.

**L<l_c=ϖ ▶ *No sound propagation, no local stress gradient with a physical meaning :***
This free-flight process has an other great consequence: be a small volume $\varpi$ of $\Omega$, it contains $n\varpi$ particles in mean; let us perturb the local pressure p in $\varpi$ and impose $p+\Delta p$ for a very short while; then $\Delta p$ shall relax with time, with a typical relaxation time $\tau$; it is straight forward to find that $\tau=\varpi^{1/3}/<v>=\varpi^{1/3}(m/\{2kT\})$. However, this perturbation vanishes as it propagates outside $\varpi$ because the standard deviation of the speed distribution of the particles is equal to the mean, *i.e.* $\delta v=<|v|>$, so that its typical extinction length is $\varpi^{1/3}$. Furthermore, as local pressure is fluctuating with time, $\Delta p$ can be considered as resulting from these fluctuations, and to be the realisation of some special configuration. It means also that sound cannot propagates, because a local periodic perturbation vanishes as it propagates.

**L<l_c=ϖ ▶ *No local thermodynamics transformation:*** The reason why sound cannot propagate in such a gas with point-like particles is that thermodynamics transformations cannot be defined (or performed at a local scale): indeed, what is an adiabatic transformation in this case on a volume smaller than $l^3$? (or an isothermal one?) Indeed, classic text books of mechanics tell clearly that one shall introduce local thermodynamics transformation in order to build the equation of sound propagation.

An other limitation for the REV can be found in some cases when two phases are present and in equilibrium; this is for instance the case of the liquid-gas equilibrium near the critical point. This happens because the rheological behaviour fluctuates on large length scale, *i.e.* neither the law pV=nRT nor an other one is valid at this point $(p_c, V_c, T_c)$. But before describing the case of different phases in equilibrium, we want to point out the importance of diffusion in the process of defining the REV. Indeed, in the case of the gas, diffusion allows to write local stress equation. So if $L<l_c=\xi$, the dynamics of the gas is the dynamics of an ensemble of independent particles submitted to forces; but when $L>l_c=\xi$, this dynamics has its own "collective" rheology which has its own laws with their own characteristics (viscosity, sound speed, diffusion coefficient).

It is worth noting however, that these "collective" laws can sometimes be determined directly from samples smaller than the REV: for instance the molecular diffusion or the molecular viscosity of a perfect gas can be determined using samples smaller than the REV; but it is because collisions occur already, even if they are not enough frequent to control the mechanics.

Before studying the case of the liquid-gas transition, let us give an other classical example when diffusion controls the REV: the electric conductor.

## *7.2. REV of electric conductor:*

An other example which allows to point out the importance of diffusion for the macroscopic law is the electric conductor: indeed, the Ohms law, U=RI (or U=ρ(l/s) j) can only be applied only on samples larger than $l_c$, where $l_c$ is the mean free path of the electron. The diffusion process can involve either electron-electron collision or electron-atom collision or electron-phonon; anyway, the electrons have to diffuse to behave as in a conductor; otherwise, the system will look like an oscilloscope.

Indeed, this is also why the diffusion coefficient D and the mobility $\mu$ of charged particles of mass m are related together via the Einstein relation:

$$\mu = qD/(k_B T) \quad (2)$$

It is because at equilibrium the particles (at density n) are forced (i) to move in the electric field so as to generate a currant density $j=\mu nqE$, *i.e.* Ohm's law, which is counterbalanced by a diffusive flux Dqdn/dx, leading to the equilibrium $\mu nqE - Dqdn/dx = 0$, whose solution is $n=n_o \exp\{-V\mu/D\}$, since $-\int Edx = V$. As this population is in *thermodynamic* equilibrium, it shall obey also $n=n_o \exp\{-qV/(k_B T)\}$. Identification of the two equations imposes Eq. (2).

### 7.3. REV in the case of phase equilibrium: the liquid-gas case:

We use the liquid-gas case as an archetype to investigate how a phase transition can modify the notion of REV. Indeed, if liquid phase exists below a given range of temperature, pressure and specific volume V, it is because the molecules of the gas exhibit *an attractive interaction at short distance* forcing the condensation into a dense liquid when the system is dense enough; (indeed the interaction becomes strongly repulsive at smaller scale, so that the system does not condense into a single point). This led Van der Waals to propose his famous equation of state whose typical isotherms are recalled in Fig. 1.

$$(p-a/V^2)(V-b) = nRT \quad (3)$$

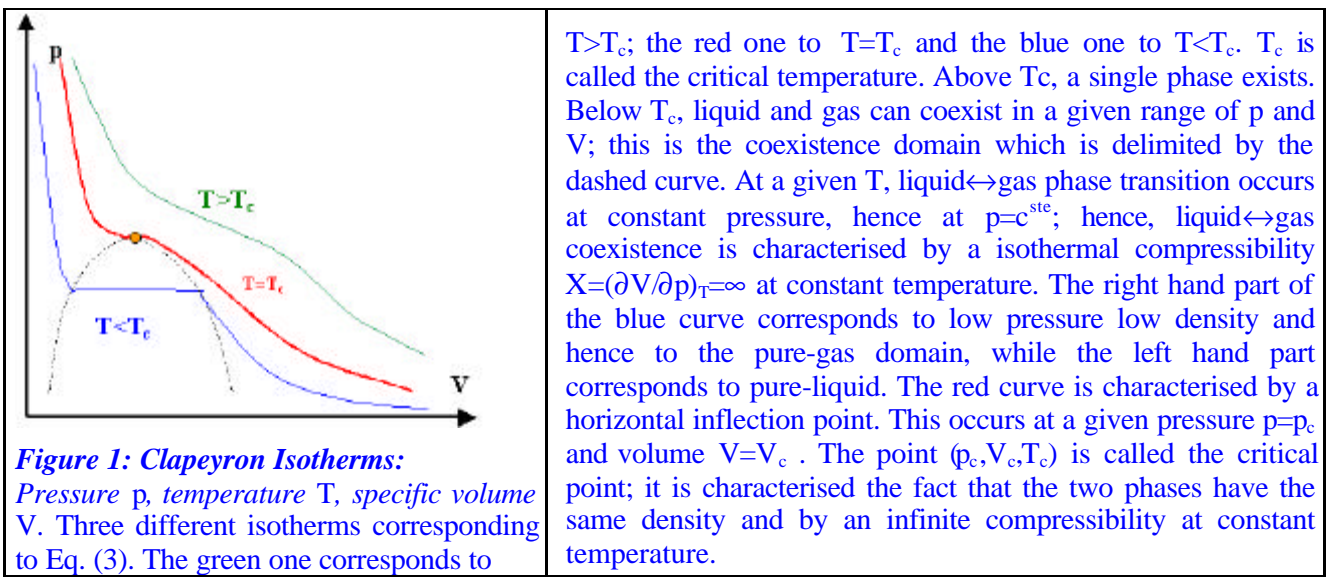

**Figure 1: Clapeyron Isotherms:** *Pressure* p, *temperature* T, *specific volume* V. Three different isotherms corresponding to Eq. (3). The green one corresponds to $T>T_c$; the red one to $T=T_c$ and the blue one to $T<T_c$. $T_c$ is called the critical temperature. Above Tc, a single phase exists. Below $T_c$, liquid and gas can coexist in a given range of p and V; this is the coexistence domain which is delimited by the dashed curve. At a given T, liquid↔gas phase transition occurs at constant pressure, hence at $p=c^{ste}$; hence, liquid↔gas coexistence is characterised by a isothermal compressibility $X=(\partial V/\partial p)_T = \infty$ at constant temperature. The right hand part of the blue curve corresponds to low pressure low density and hence to the pure-gas domain, while the left hand part corresponds to pure-liquid. The red curve is characterised by a horizontal inflection point. This occurs at a given pressure $p=p_c$ and volume $V=V_c$. The point $(p_c, V_c, T_c)$ is called the critical point; it is characterised the fact that the two phases have the same density and by an infinite compressibility at constant temperature.

Hence, phase transition occurs at small enough temperature ($T<T_c$) due to the competition between disorder and energetic balance : as the molecules attract each other they try to condense; this occurs at high enough density since, if their mean distance is too large they do not see each other sufficiently to condense. Anyhow, the transformation of one phase into the other produces (gas→liquid) or needs (liquid→gas) energy $\Delta H=0$. This is the sub-critical domain.

At larger temperature (T>$T_c$), the interaction between the molecules is smaller than the mean kinetic energy, so that the interaction is not large enough to allow condensation anymore and the molecules remain in a gas phase. This is the super-critical domain.

So it exists a temperature $T_c$ for which the liquid phase starts to exist. This occurs at a definite value of pressure $p_c$ and volume $V_c$. The gas-liquid transition is just entropic; and the two phases are difficult to distinguish; this generates fluctuations at large length scale and on large time scale, which are known as critical fluctuations and critical slowing down.

Before looking at the case T=$T_c$, let us recall the behaviours far away the coexistence domain, then we will consider the case of phase coexistence and lastly the case near ($T_c$, $p_c$,$V_c$).

***Far from ($T_c$, $p_c$,$V_c$) and not in the coexistence domain :*** In this case, the system is made of a single phase, which is either a gas or a liquid, and the densities of the liquid ($\rho_l$) and of the gas ($\rho_g$) are quite different. $\Delta\rho=\rho_1-\rho_2$ is called the order parameter.

*A unique Gas phase :* As it is far from the critical point, the gas has a small density; it is a gas of rather independent molecules which interact via collisions. However, there is some probability of getting clusters of molecules , *i.e.* pairs, triads, etc., due to the existence of the short range interaction; but their proportions remain small and decrease fast with the cluster complexity (and when V→0), because collisions are rare.

It results from the above analysis that one can apply continuum mechanics on volumes larger than $l_c$, with $l_c$ given by Eq. (1).

At a given temperature T<<$T_c$, when the gas density increases, larger clusters appear more frequently due to collisions; but they are still in little proportion because they are unstable compared to the pure liquid phase; this is due to a surface effect which makes a larger droplet more stable than a smaller one so that liquid phase appears before small clusters are numerous below $T_c$. This leads to introduce a critical radius $R_c$ for the coalescence of large drop of liquid. So, even at the limit of liquid condensation, the clustering process does not perturb too much the mechanics of the perfect gas when T is far from $T_c$, and the REV remains of the size given by Eq. (1).

*A unique Liquid phase :* In the liquid phase, the density is rather large and merely constant, if one excepts the region near the critical point. It means that the mean free path $l_c$ is the molecule size D, $l_c \approx D$. Furthermore, the molecules organised themselves at short distance; this generates some order and structures the first neighbours; but this structure becomes random a little further, *i.e.* for second neighbours already. This randomness allows also to determine the constitutive law from averaging of different local configurations; the size of the typical configurations define the REV. Owing to this, the REV size turns out to be 2 or 3 D about ($\xi$<3D).

***In the coexistence domain, but far from $T_c$ :*** The two phases are in equilibrium and exchange molecules. But the characteristics of each phase remain quite different, well defined and constant at a given temperature whatever the relative concentration of the two phases. Hence both have the REV defined in the preceding subsection, *i.e.* the gas phase has a REV given by Eq. (1), the liquid phase a REV of about $(2or3D)^3$.

under thermodynamics equilibrium the two phases should be continuous and form a single continuum each separated by a smooth interface, either a sphere or a plane. In this case the mechanics problem, such as the one of sound propagation, can be handled simply, by writing the boundary conditions at interface. However dynamics and mixing can generate more complicated geometries and complex interfaces such as homogeneous or non homogeneous dispersion of droplets or of bubbles of different size; it can be more complex and continuous, with one of the phase forming the porous material, the other phase occupying its interior. In this case the problem of sound propagation can generate scattering, diffusion and perhaps localisation for certain frequency range.

*Near $T_c$ :* Fig. 1 shows that at $T_c$ both phases have the same density and are hyper-compressible; the energy of transformation from one phase to the other one vanishes and the problem becomes only controlled by entropy. It results from this that both phases fluctuate in time and space, transforming permanently, one in the other and conversely. But fluctuations can also propagate, because the system is soft and do not dissipate. This produces fluctuations that do not disappear on large volume. Furthermore, once slightly perturbed, the system needs an infinite time to come back to equilibrium . These phenomena are known as critical fluctuations and critical slowing-down. This imposes that the REV is infinite at the critical point. As a consequence of the slowing down, one finds also that the speed of sound tends to 0. However this requires to be exactly at $(T_c, p_c, V_c)$. A way to approach and model these phenomena is to apply ideas developed in the Renormalisation Group (RG) theory.

The idea which underlies RG is the following: In the vicinity of the critical point the system develops classic behaviour on length scales sufficiently large; one can then define a finite REV. However, at the critical point the system develops fluctuations at all length scales. These fluctuations are induced by the "continuous medium" itself which fluctuates spontaneously in an anomalous way because the local fluctuations at a given scale propagate to adjacent volumes due to mechanical coupling; hence it forces fluctuations at larger scales; damping of these fluctuations due to energetic consideration is not possible because all states have similar energy. Of course, this requires also to identify the physical property which fluctuates; this defines the order parameter. (In the case of the liquid gas transition it is the density).

As these anomalous fluctuations come from the continuum limit due to coupling between scales, one may think to handle the problem via a change of scale. This is the RG procedure. The first step, i.e. step 0, requires to identify the physical property which fluctuates and the order parameter. Then the RG procedure consists in integrating the system by averaging it on smaller length scale, and averaging the fluctuations at small length scale. So one starts with a discrete system of interacting particles , with a mean distance of particle a and an interaction strength $\langle J_{ij}\rangle$. One integrates the Gibbs function from a→sa . This is the first operation of RG. The second operation of RG consists in rescaling the length scale in order to be able to compare the new system with the older one. However, this rescaling forces also to renormalise the coupling in order that the previous system becomes equivalent to the new one; this is the third operation of RG. So, this procedure shall impose that the physics remain the same at a given legth scale $L\gg a_1$; it necessitates to multiply the size of the microscopic objects by s ($a_1 \rightarrow a_2=sa_1$), then to reduce the length scale and endly to renormalise the interaction strength ($<J_{ij}>_1 \rightarrow <J_{ij}>_2=A_J<J_{ij}>_1$).

On the other hand changing the interaction strength ($<J_{ij}>_1 \rightarrow <J_{ij}>_2=A_J<J_{ij}>_1$) without changing the space length allows to predict the fluctuations at a size sL instead of L. So this procedure allows changing of scale and can be iterated. Let us apply it on a physical property M; it is $M_1$ at scale $a_1$; it becomes $M_2=A_M M_1$, ..., $M_{n+1}=A_M^n M_1$ ,... at scales $a_2=sa_1$, ..., $a_{n+1}=s^n a_1$ ,...respectively. This implies that M scales as $M/M_1 \propto (L/a_1)^\alpha$, with $\alpha=\ln_s A_M$. This is why physical properties obeys scaling laws with size. It is due to the internal dilation symmetry the system shall exhibit just at the critical point.

The fixed points of the RG transformation are critical points when the exponent are not classic, *i.e.* when they are not those obtained from dimensional analysis . In this case they are unstable for at least one parameter which is the temperature $(T-T_c)$. When they are unstable for two parameters they are called tri-critical points…. Of course a critical point is an unstable point, since classic behaviour is recovered at large length scale when $(T-T_c)\neq 0$.

The scaling laws of liquid-gas transitions are reported in Table 1, and the values of the critical exponent in Table 2.

In the vicinity of the critical point, *i.e.* $T\neq T_c$, the dilation symmetry is broken. However dilation symmetry remains approximately valid on small volumes: as the physics is governed by the same local interactions as previously, the system generates a lot of states with little energy difference on small volumes; and their statistics shall obey the same distribution as those at the critical point. However, the anomalous fluctuations that were also generated at $T=T_c$ on larger length scale, are no more possible at $T\neq T_c$ because they generate too much energy difference at this larger scale. So, the system continues generating a lot of fluctuations on small scales; but it exists a maximum length, called the correlation length $\xi$, above which the relative fluctuations diminish normally, *i.e.* above which the central limit theorem applies. $\xi$ depends on temperature as a power law, *i.e.* $\xi=(T-T_c)^{-\nu}$.

| property | scaling | property | scaling | property | scaling |
|---|---|---|---|---|---|
| density | $\Delta\rho/\rho_c = B\,|\varepsilon|^\beta$ | Thermal diffusion | $D_T \propto D_{To}\,|\varepsilon|^{\gamma-\Psi}$ | Surface tension | $\sigma \propto \sigma_o\,|\varepsilon|^\mu$ |
| Isothermal compressibility | $K_T \propto |\varepsilon|^{-\gamma}$ | Specific heat at constant pressure | $C_p \propto C_{po}\,|\varepsilon|^{-\gamma}$ | Thermal conductivity | $\Lambda \propto \Lambda_o\,|\varepsilon|^{-\Psi}$ |
| Specific heat at constant volume | $C_v \propto C_{vo}\,|\varepsilon|^{-\alpha}$ | Correlation length | $\xi \propto \xi_o\,|\varepsilon|^{-\nu}$ | Speed of sound | $c \propto c_{so}\,|\varepsilon|^{\alpha/2}$ |

*Table 1: Scaling laws and critical exponents. Here $\varepsilon$ stands for $\varepsilon=(T-T_c)/T_c$.*

Furthermore, as all the physical properties are controlled by a single parameter near $T_c$, *i.e.* $T-T_c$, (in the case of critical point), once one knows the size of the representative volume $\xi$, one can determine $(T-T_c)/T_c$ to which it corresponds by using the scaling .

From this value one can determine the other physical properties using the relations in Tables 1 & 2.

For instance, from Table 1, one sees that the speed of sound c tends to 0 as $[(T-T_c)/T_c]^{\alpha/2}$ when $(T-T_c)/T_c$ tends to 0; this indicates the divergence of the REV; however, the exponent on c is small, *i.e.* $\alpha/2 \approx 0.06$, so that it is hardly seen before $(T-T_c)/T_c < 10^{-4}$. On the other hand, the correlation length scales as: $\xi=\xi_o(\Delta T/T_c)^{-\nu}$, with $\nu=0.63$ and $\xi_o=1.5$ to 2 Å for the liquid gas transition. So $\xi > 500$ Å when $(T-T_c)/T_c < 10^{-4}$.

| Critical exponent | α | β | ν | ψ | μ | γ |
|---|---|---|---|---|---|---|
| classic value | 0 | 1/2 | 1/2 | - | 3/2 | 1 |
| RG value (d=3) | 0.11 | 0.325 | 0.63 | 0.57 | 1.26 | 1.241 |

*Table 2: Critical exponents. RG stands for renormalisation group performed on 3d systems*

As a conclusion of the mechanics of a liquid-gas problem, they are two length scales involved in defining the REV. The first one is the mean free path $l_c$ between collisions; it comes from the diffusion process since the particles have to interact to define local thermodynamic equilibrium. The second limitation is that the thermodynamics limit can be taken on a small volume: in some cases when the difference of energy in between the two phases is extremely small and tends to 0, the system is governed by entropy only, so that it fluctuates on all distances. This last case occurs in a phase transition of the second order kind, whose typical examples are the liquid-gas transition at the critical point or the fero-para magnetism transition. At $T_c$ the REV becomes infinite; near $T_c$ it diverges as $\xi=\xi_o(\Delta T/T_c)^{-\nu}$.

### 7.4. The REV in the case of crystals and poly-crystals :

In the case of a perfect crystal, the molecules are in interaction with nearest neighbours, the medium is homogeneous and the elastic properties are defined mainly from the elementary cell; hence this one defines the REV. However the REV could be larger if the interaction with second and third nearest neighbours have to be introduced.

After a while, when the crystal is deformed sufficiently, plastic deformation is generated, which is linked to the generation of dislocations. As far as dislocations do not interact together the true REV shall contain a dislocation in mean so that it can be large. However, the interaction between dislocation and the crystal results essentially in the perturbation of the elastic field, even at the length scale of the elementary cell. And the elastic field obeys linear differential equations, whose solutions are additive. So one can treat the system using mean field as a superposition of coupled elementary cells, some of them being perturbed by a dislocation. The density of dislocations and the dislocation generation is controlled by the elastic strain so that the REV remains of the cell size.

When deformation gets larger, dislocations start interacting together, leading to the creation or the anihilation of dislocations. The REV shall contain such interaction; this requires the REV size to be equal to the typical distance of dislocation-dislocation interaction.

If the systems is formed of poly-crystals, the interaction between the different crystals becomes the new typical REV size, and so on. So one sees that a successive series of length scales shall be introduced as the system becomes more and more complex. When the crystals exhibit important anisotropic response, the stress in the polycrystal can be concentrated on chains of crystals as in the case of granular materials; but these chains do not propagate further than 10-crystal diameter [35].

If one imagines that this series of lengthscale is generated by some self-similar process, this builds up a system whose complexity increases with size as a self-organised critical system. However, in the case when the hierarchy of size is well separated, for instance by few orders of magnitude at each step, such a self-organised process can be difficult to identify, and one can conclude to the presence of successive distinct length scales without connection instead of a hierarchical determinist structure.

## 8. Conclusion:
### The REV of a granular medium from papers at Powders & Grains 2001:

In §-7.1-§-7.4, it has been recalled that defining the mechanical REV of a medium needs to define two length scales. (i) The first one, $l_c$, corresponds to the distance of interaction; (ii) the second one, $\xi$, is the size of the elementary volume on which one can apply the "thermodynamics" limit. Point (ii) implies two rules: (a) if the system is larger than the REV, its mean behaviour remains the same if its size is doubled; (b) the fluctuations the mechanical response exhibits shall obey the central limit theorem.

*The REV from fluctuations :* Most experiments or simulations try to identify the REV to the typical size of the system above which fluctuations become negligible. According to point (b), this is not correct and the REV can be much smaller. So, a right way to determine the length $\xi$ of the REV is to analyse the relative fluctuations of the q *vs.* $\varepsilon_1$ curve : as q is a stress and as it is the sum of random forces $F_i=q_i\delta S$, whose number shall grow as $L^{d-1}/\delta S=(L/\xi)^{d-1}$, one expects that $\delta q/q$ shall vary as $(\xi/L)^{(d-1)/2}$, *i.e.* as $\xi/L$ above $\xi$ in 3d and as $(\xi/L)^{1/2}$ in 2d. Unfortunately, most DEM simulations do not look at the speed of convergence of the curve (q/$\sigma_3$-*vs.*-$\varepsilon_1$) with the size L/D of the system so that it is not yet possible to identify $\xi$ from this method.

Anyhow, this method can be used indirectly, *via* the study of the fluctuations for a given L/D. This can be done on experiments: typically, the experimental fluctuations of the q/$\sigma_3$ *vs.* $\varepsilon_1$ curve is about 1%-3% for a sample size to grain diameter ratio L/D=100. Assuming that the relative standard deviation of the stress field in the REV is 1, one finds that experimental fluctuations correspond to $\xi_{REV}$ = D-to-3D .

The same method can be applied to DEM simulations and one finds the same typical value, *i.e.* $\xi_{REV}$ = D-to-3D.

*The REV from avalanches :* It is worth noting however that avalanche experiments have concluded to a much larger REV size, since it is about $\xi/D= \delta\theta=40$ [10,12,13].

***REV from acoustics:*** It is known from long that sound propagates in granular media; this propagation is used in test labs to determine the elastic constant of granular samples containing few millions grains; it is also used on Earth to perform acoustic imaging... Indeed, Jia's work, recalled in §-6, has re-demonstrated this point; it has also studied the diffusion limit, in the case of acoustic waves of small wavelength and high frequency. So this demonstrates that a REV can be defined on cells of few millions grain, and that the mechanical characteristics do not fluctuate so that the real REV is likely smaller.

***REV & Finite Element calculations:*** Also macroscopic stress FE calculations using classic rheology of granular materials agree with experimental findings in piles, *cf.* §-5.2. and with DEM, at least within the experimental uncertainty. In particular, FE calculations allow to predict the dip of stress in conic piles built from a point source and the sensitivity of the stress distribution to the building process. This means that macroscopic mechanic behaviour is relatively well defined and that macroscopic averaging is possible.

***REV & Discrete Element calculations:*** The existence of well defined rheological rules is confirmed by DEM simulations of Combe & Roux, *i.e.* §-1, as far as grains exhibits solid friction; this seems no more true when grains are frictionless, see §-1, (but this is a problem with little physical meaning). A possible explanation for this last result is that the internal rheology of the sample does not contain any energy term so that it is controlled by entropy, hence leading to critical behaviour.

Anyway these results are in agreement with DEM simulations of Radjai on grains with friction, see §-2. Radjai has tented to discriminate between two sub-networks of strong and weak contact and forces. The first one supports deviatoric stress the other one not. However, the contact status, *i.e.* defined as the belonging to one sub-network or the other, is found to be sensitive to calculus precision, to a tiny change of some contact force and to evolve fast with deformation. So we have concluded that the two sub-networks are in "thermodynamics" equilibrium and can be considered as forming a single phase with two constituents. This is true till the medium contains some larger grains; in that case, the larger grain condense some definite part of the two sub-networks, forming then a second phase that does not mix.

Anyhow, it turns out that the REV is always much smaller than the systems used for these Discrete simulations, except when grains are frictionless for which an anomalous scaling of the fluctuations with sample size has been observed. Estimate of REV size can be done from noise on q *vs.* $\varepsilon_1$ curves; it leads to $\xi_{REV}$ = D-to-3D.

***REV from the experimental force network:*** Another way to determine the REV is from studying the local structure of the force network, via correlations. This can be done via Dantu-like experiments with photoelastic disks either in triaxial- or in Couette- geometry (§-4), or via DEM simulations (§-2), or with experiments specially dedicated to observe the "linear" propagation of stress (§-5.a). Indeed forces seem to propagate along linear chains; but the correlation length of these chains is small, *i.e.* $\xi$=2 to 5 grains, in all classic cases. Cases where longer chains are observed are rare and mainly pathologic: they are observed either (i) when the medium is ordered and submitted to a point like perturbing force (§-5.a), or (ii) in Couette geometry when density of grains becomes small, so that contact with boundary occurs intermittently. Furthermore, the length of a chain does not overpass 10-grain diameter in both cases. Both cases are pathologic, because case (i) does not correspond to the characteristics of an ideal random granular medium and because case (ii) can be due to imperfection of the experiment, as the roughness of the boundary, or to finite size effect, as in sand avalanching.

However, case (ii) can also be the signature of a critical behaviour. If so it would be characterised by $\rho_c$, *i.e.* the smallest possible density, and by $\sigma \rightarrow 0$. Indeed, these conditions are not the usual ones for working. However it might be found in some cases, for instance during the gravitational condensation that compacts asteroid, planet and stars during their formation, if electrostatic forces are not too important.

It is worth noting that similar condensation of stress in linear channels of monocrystals has been found also in elastic polycrystals with monocrystals with non isotropic elastic response [35].

***REV size as a problem of Force diffusion:*** Indeed, we have pointed out the importance of the mean free path $l_c$ in order to define the REV of a gas of particles. In this sub-section, we want to state the parallel between $l_c$ and the length of the force chains. Let us first remark that each particle of a gas transport a momentum mv ; but as it moves with speed v, the number of collisions the particle makes per unit time depends on its speed v so that it wear a mean momentum transfer **T** per unit of time equal to **T** =mv². This mean transfer of momentum per unit of time is equivalent to a force per unit area area, or to a pressure. This is indeed well known, and results in the law of perfect gas pV=nRT .

Anyway, as **T** is the analogue of a force, the gas can be viewed as a random distribution of T vector. So, one can try to compare it to the force network in the granular medium. So let us consider that each gas particle transport some quantity **T** called transfer of momentum. **T** is a vector of amplitude **T** =mv². Its statistics follows p(**T**)=exp{-**T**/(k$_B$T)}. At some time, the T vectors are distributed randomly. Each **T** propagates along a line whose typical length is $l_c$. Then it diffuses in an other direction with a different strength **T'**. Let us now compare that to a granular medium.

First the force network of a granular medium exhibits also an exponential distribution p(**F**)=exp{-**F**/F$_o$}. The parallel may then "explain" simply why the granular force obeys an exponential distribution: in gas it is due to the principle of maximum disorder applied with some constrain to preserve; in granular matter, [36] has demonstrated that the same principles do apply and explain the distribution p(**F**)=exp{-**F**/F$_o$}, see [36]. Then this analogy supports the validity of an approach "à la Boltzmann"[36].

Also the forces propagate along "linear" chains before diffusing in granular matter. The typical length of these chains is $l_F$. According to the analogy, $l_F$ has to be compared to the mean free path of a gas, *i.e.* $l_c$ .It turns out that $l_F$ is in general much smaller than the $l_c$ of a dilute gas, *i.e.* it is at most 2- to 3- grain diameter. Furthermore, it is possible that the observed value of $l_F$ is increased because the force network of a granular material shall satisfy two requirements instead of one in the case of a gas: *requirement 1:* necessity of having a broad distribution of forces; *requirement 2:* conditions of local equilibrium. Indeed, requirement 2 is not needed in the gas case; if one tries to apply these two requirements to the local equilibrium of disk, one gets the following: as a typical disk has 4 contacts about, the first condition imposes that halve the contacts of the grain wear small forces, while the other halve wear large forces. Second condition imposes that large forces are in equilibrium, which imposes then that large forces are aligned when the mean number of contacts is 4. Nevertheless, concentration of stress along linear chains is not the only property of rigid grain assembly, since it has been found experimentally and explained numerically also in poly-crystals[35].

So, if we pursue the analogy with gas, as the chain length $l_F$ is also the length that ensures the diffusion of the force, $l_F$ shall correspond to the REV size $\xi$, in most cases, since $l_c$ is also the REV of the gas most often. It would mean that the REV is quite small in most granular experiment and corresponds always to 2- or 3- grain diameter. *This value $\xi$=D-to-3 D is then the typical value to which each experiment or computer simulation leads.* This explains why granular materials exhibit well defined macroscopic rheological law, which we hope can be simplified, as proposed in [37]. But this is an other problem.

*REV and stress propagation along lines:* In the last sub-section, we have related the diffusion length $l_F$ of the forces to the REV size $\xi$. Following this analogy, one can ask also whether the REV size can be increased by increasing $l_F$. If so, a way to increase $l_F$ would be perhaps to run the system into a plastic state. Indeed in this case, plasticity will impose a relation between the two stress components, leading to hyperbolic equation whose solution gives that stress should propagate linearly as a ray; so this should be able to impose that the forces propagate linearly and that $l_F$ increase. It is surprising that such a behaviour has not been observed yet in recent experiments.

*REV and natural conditions:* This paper has investigated the REV size obtained in lab experiments. It has been found in this case that the rheologic behaviour is well defined in most cases, that it does not fluctuate and that the REV is smaller than $\xi$<3D. This is true because experiments in lab conditions exhibit well defined condition, and because samples were homogeneous. However, it is known also that the mechanical behaviour of granular matter is highly non linear and exhibits irreversibility, which leads to story-dependent behaviours. So this could lead that natural soil exhibits less reliable behaviours, because they have been submitted to diverse history, to heterogeneous loading and because the past loading can be smaller or larger than the future one at will.

*REV in stick-slip conditions:* As a conclusion, it is worth quoting recent experimental study [38] which may contradict the universality of the above conclusion on the smallness of the REV size. Indeed, in this work [38], the stick-slip mechanism has been studied in granular samples made of glass spheres and submitted to uniaxial axi-symmetric compression at constant rate of deformation. It has been found that the experimental results were different in small samples, *i.e.* V=(200)$^3$ grains, and in large samples, *i.e.* V>(1000)$^3$ grains; for instance, one observes the passage from a regime of random non correlated stick-slips with random amplitude to a regime of quasi-periodicity. This shows that the REV can be much larger than $\xi$=200D in some cases. Indeed, this behaviour can result from a cross-over between small erratic response with large dispersion to quasi-periodic regime made of events of similar size, as it is already observed for avalanches; and its interpretation can be similar to the avalanche one, defining the REV length $\xi$=200D-500D. However this value of the REV size has to be taken with caution because the experiments have demonstrated also that the stick-slips in both kinds of sample are produced by a dynamic instability related to a weakening of the mechanical response when the strain rate is increased. The change of response could be due to some dynamical effect which depends on the sample size, without needing the introduction of a large REV.

*Acknowledgements:* CNES is thanked for partial funding.

## Appendix: SOC , and classical criticality

First and Second order phase transition have been described briefly in §-7. It has been recalled that second order phase transition exhibits critical fluctuations which imposes a diverging REV, imposing a slowing down dynamics, the existence of critical scaling defined by critical exponents. In particular, the relative fluctuations does not decrease with the system size according to the central limit theorem in such systems, but much slower, with a power law. In general critical behaviours are not observed "easily": for instance, in the case of the liquid-gas transition it occurs just at $(T_c,p_c,V_c)$, which is just a point in a 3d phase space; so the probability that it is achieved spontaneously is quite small and its observation requires a careful research.

On the contrary, it has been proposed by Bak, Tang & Wiesenfeld (BTW) [4] a simple computer algorithm which is running always in critical conditions and which produces noise with critical fluctuations. Its archetype is a sandpile slope and the flow it generates when adding grains: be $(x_i,y_j,z_k)$ a system of discrete coordinates forming a regular simple cubic network; these points can be occupied by grains; be a finite horizontal (x,y) surface at z=0 with a boundary $x\pm y=\pm L_o$ ; build a 3d pile from a point source (0,0,h) , $L_o$<h, by letting fall grains vertically, one by one, in such a way that complete equilibrium is obtained before letting fall the next grain. The flowing rules of the grains on the pile surface are as follows: they are stable if the height in between two adjacent columns is smaller than $\delta h$; but if it is larger the higher 2 grains of the higher column flows on the two adjacent column, one on each column. After running the algorithm during a while, an intermittent flow is generated outside the basis surface. This flow is made of a series of avalanches having a broad distribution of size N, whose probability scales as $P(N)=N^{-\alpha}$. Here N is the number of grains contained in the avalanche.

This algorithm was an important result because it was the first example of a system that was working spontaneously and always under critical conditions; this is why it was called Self Organised Criticality (SOC). As the algorithm was aimed at simulating sand avalanches, one was thinking to be able to observe it there. This is not the case [10-12] except for small piles [13].

Looking for SOC behaviour means looking for a generalised-critical trend. So, it is not as a single critical point which is difficult to observe. It shall be found easily due to the "self organisation". So, the existence of few critical points in the phase space is not equivalent to SOC except if one would able to force the system working always under critical condition.

As in any problem of criticality, SOC implies the existence of a hierarchy of structures and of lengths. Hence, if granular matter is the archetype of SOC, no good average behaviour can ever be defined. On the contrary, when classic average can be defined for almost all working points, with Gaussian fluctuations, it can happen, however, that the behaviour becomes more complex for some set of parameters. This is for instance the case for the liquid-gas phase transition for which liquid and gas have well defined behaviours except at the critical point.

Owing to this the real question which is addressed in the present paper is : is any experiment on granular matter always sensitive to uncontrolled details,…?

The electronic arXiv.org version of this paper has been settled during a stay at the Kavli Institute of Theoretical Physics of the University of California at Santa Barbara (KITP-UCSB), in june 2005, supported in part by the National Science Fundation under Grant n° PHY99-07949.

*Poudres & Grains* can be found at :
http://www.mssmat.ecp.fr/rubrique.php3?id_rubrique=402